\documentclass[aps, reprint, nofootinbib, showpacs, floatfix]{revtex4-1}

\usepackage{amsmath}
\usepackage{amsfonts}
\usepackage{amssymb}
\usepackage{graphicx, rotating}
\usepackage{epstopdf}
\usepackage{epsfig}
\usepackage{latexsym}
\usepackage{graphicx}
\usepackage{color}
\usepackage{amsmath,amssymb}
\usepackage{multirow}
\usepackage{hhline}
\usepackage{array,makecell}

\usepackage{slashed}
\usepackage{hyperref}
\hypersetup{colorlinks, citecolor=bluscuro, linkcolor=black, urlcolor=bluscuro}
\definecolor{rossos}{cmyk}{0,1,1,0.55}
\definecolor{bluscuro}{rgb}{0.15, 0.2, .85}
\definecolor{bluchiaro}{cmyk}{1,.3,0.,0.1}

\newcommand{\eq}[1]{Eq.~(\ref{#1})}
\newcommand{\lag}{\mathcal{L}}

\newcommand{\nn}{\nonumber}

\newcommand{\be}{\begin{equation}}
\newcommand{\ee}{\end{equation}}
\newcommand{\bea}{\begin{eqnarray}}
\newcommand{\eea}{\end{eqnarray}}
\newcommand{\bc}{\begin{center}}
\newcommand{\ec}{\end{center}}

\newcommand{\op}{\,\mathcal{O}}

\begin{document}

\widetext

\title{Higgs Couplings without the Higgs}
\author{Brian Henning}
\affiliation{D\'epartment de Physique Th\'eorique, Universit\'e de Gen\`eve,
24 quai Ernest-Ansermet, 1211 Gen\`eve 4, Switzerland}
\author{Davide Lombardo}
\affiliation{D\'epartment de Physique Th\'eorique, Universit\'e de Gen\`eve,
24 quai Ernest-Ansermet, 1211 Gen\`eve 4, Switzerland}
\author{Marc Riembau}
\affiliation{D\'epartment de Physique Th\'eorique, Universit\'e de Gen\`eve,
24 quai Ernest-Ansermet, 1211 Gen\`eve 4, Switzerland}
\author{Francesco Riva}
\affiliation{D\'epartment de Physique Th\'eorique, Universit\'e de Gen\`eve,
24 quai Ernest-Ansermet, 1211 Gen\`eve 4, Switzerland}

%\date{\today}

\begin{abstract}
\noindent

The measurement of Higgs couplings constitute an important part of present Standard Model precision tests at colliders. 
We show that modifications of Higgs couplings induce energy-growing effects in specific amplitudes involving longitudinally polarized vector bosons, and we initiate a novel program to study these very modifications of Higgs couplings off-shell and at high-energy, rather than on the Higgs resonance.  Our analysis suggests that these channels are complementary and, at times, competitive with familiar on-shell measurements; moreover, they offer endless opportunities for refinements and improvements.
\end{abstract}

%\pacs{yyy}

\maketitle
\medskip

\section{Introduction}
The precise measurement of the Higgs boson couplings to other Standard Model (SM) particles is an unquestionable priority in the future of particle physics.
These measurements are important probes for our understanding of a relatively poorly measured sector of the SM; at the same time they offer a window into heavy dynamics Beyond the Standard Model~(BSM). Indeed, it is well-known that the exchange of heavy states (with masses beyond the direct collider reach)  leaves imprints  in low-energy experiments, in a way that is systematically captured by  an Effective Field Theory~(EFT).

 There are a number of similar ways in which one can parametrize modifications of Higgs couplings~(HC): via  partial widths $\kappa_i^2=\Gamma_{h\to ii}/\Gamma^{\textrm{SM}}_{h\to ii}$~\cite{Heinemeyer:2013tqa}, via  Lagrangian couplings in the unitary gauge $g_{hii}$~\cite{Gupta:2014rxa,deFlorian:2016spz}, via pseudo observables~\cite{Gonzalez-Alonso:2014eva}, or via the effective field theory $\lag=\sum_i c_i\op_i/\Lambda^2$, consisting of dimension-6 operators~\cite{Grzadkowski:2010es,deFlorian:2016spz}. %Operators of the form $|H|^2\op_{\textrm{SM}}$, with $\op_{\textrm{SM}}$ a dimension-4 SM operator,
 In particular, the operators
\begin{gather}
\op_r=|H|^2\partial_\mu H^\dagger\partial^\mu H\,\quad
\op_{y_\psi}=Y_\psi|H|^2\psi_L H \psi_R \nn\\
\op_{BB}=g^{\prime\,2}|H|^2 B_{\mu\nu}B^{\mu\nu}\quad
\op_{WW}=g^2|H|^2 W^a_{\mu\nu}W^{a\,\mu\nu}\nn\\
\op_{GG}=g_s^2|H|^2 G^a_{\mu\nu}G^{a\,\mu\nu}\quad \op_6=|H|^6\label{eq:ops}
\end{gather}
with $Y_\psi$  the Yukawa for fermion $\psi$, can be put in simple correspondence with the $\kappa$s, as they modify single-Higgs processes without inducing other electroweak symmetry breaking effects.
%\com{Include Brian discussion on linear vs non-linear vs kappas}

The well-established method for testing HC  is, of course, to  measure processes in which a Higgs boson is produced on-shell.

\begin{table}[h]
	\begin{center}
		\begin{tabular}{ >{\centering\arraybackslash}m{0.7cm}| >{\centering\arraybackslash}m{1cm}|>{\centering\arraybackslash}m{2cm}|>{\centering\arraybackslash}m{2.5cm}|>{\centering\arraybackslash}m{1cm}}
			%\hhline{~|--|}
			\multicolumn{2}{ c| }{} &  {HC} & {HwH}&Growth  \\ 
			%\hhline{-|=||=|}
			\hline
			{$\kappa_{t}$} &{$\op_{y_t}$}& {\resizebox{2cm}{1.2cm}{
					\includegraphics[width=0.3\textwidth]{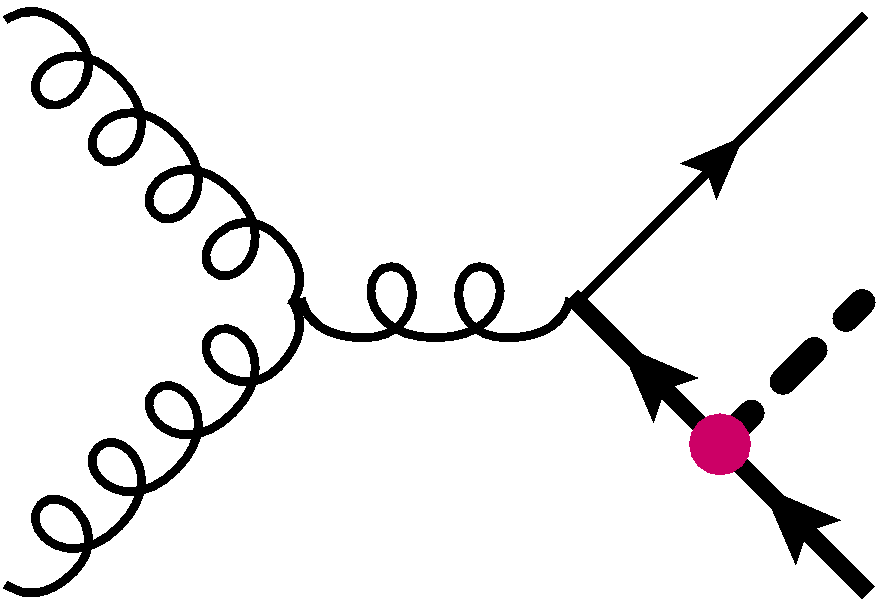}
			}} & {\resizebox{2.3cm}{1.8cm}{
					\includegraphics[width=0.3\textwidth]{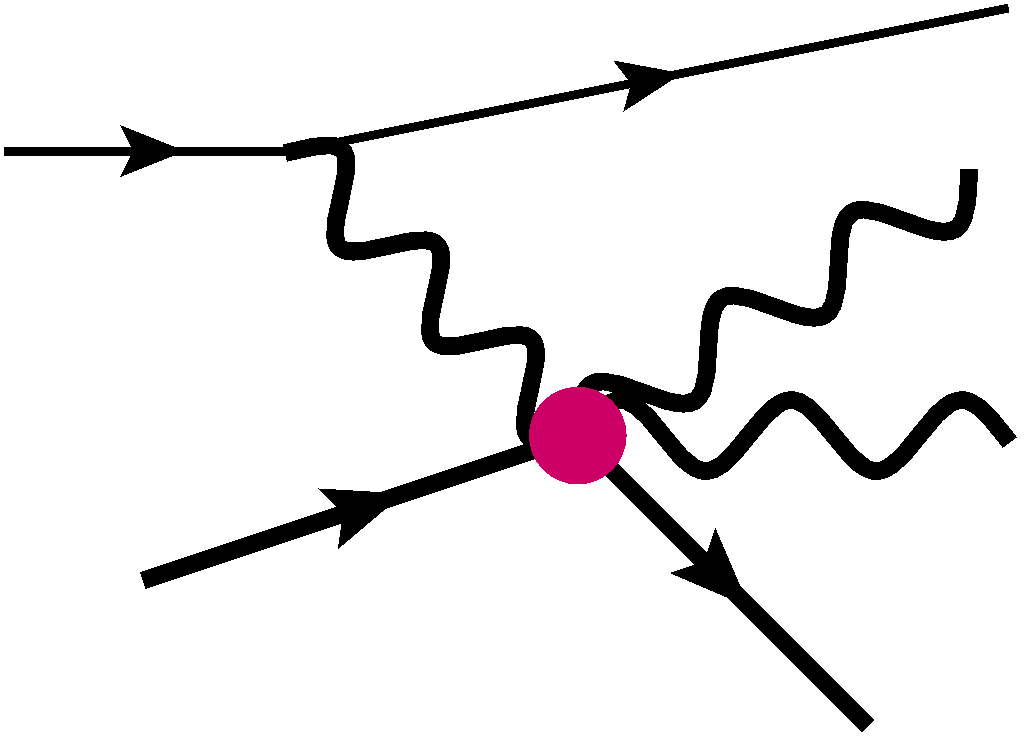}
			}}&$\sim\frac{E^2}{\Lambda^2}$ \\ \cline{1-5}
			{$\kappa_\lambda$} &{$\op_6$} & {\resizebox{2cm}{1cm}{%
					\includegraphics[width=0.3\textwidth]{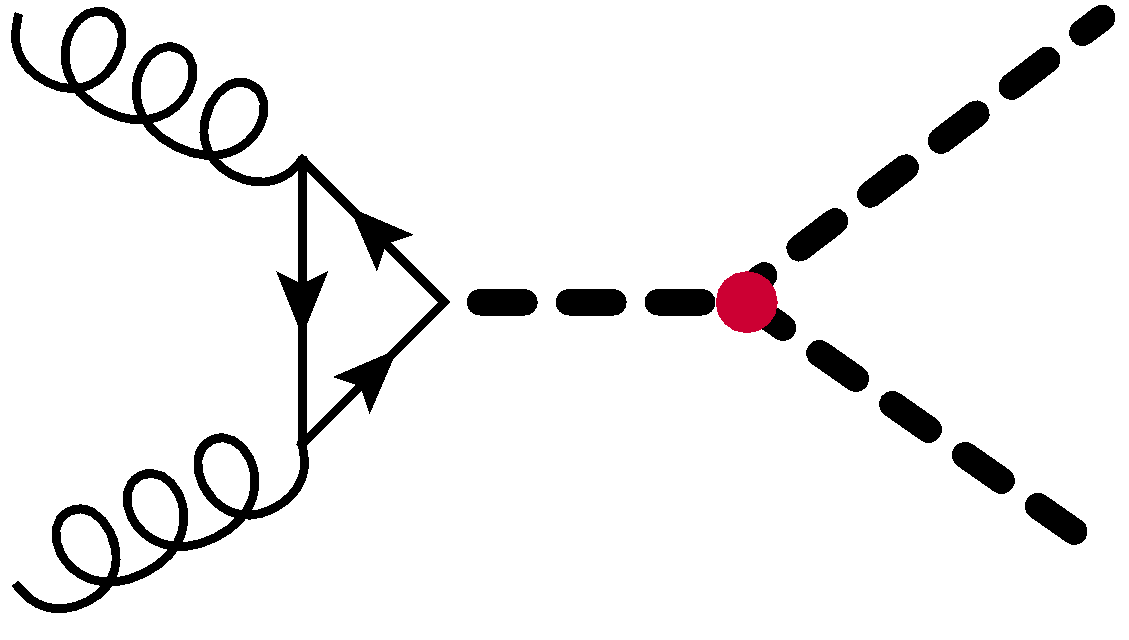}
			}} &  {\resizebox{2cm}{1.8cm}{
					\includegraphics[width=0.3\textwidth]{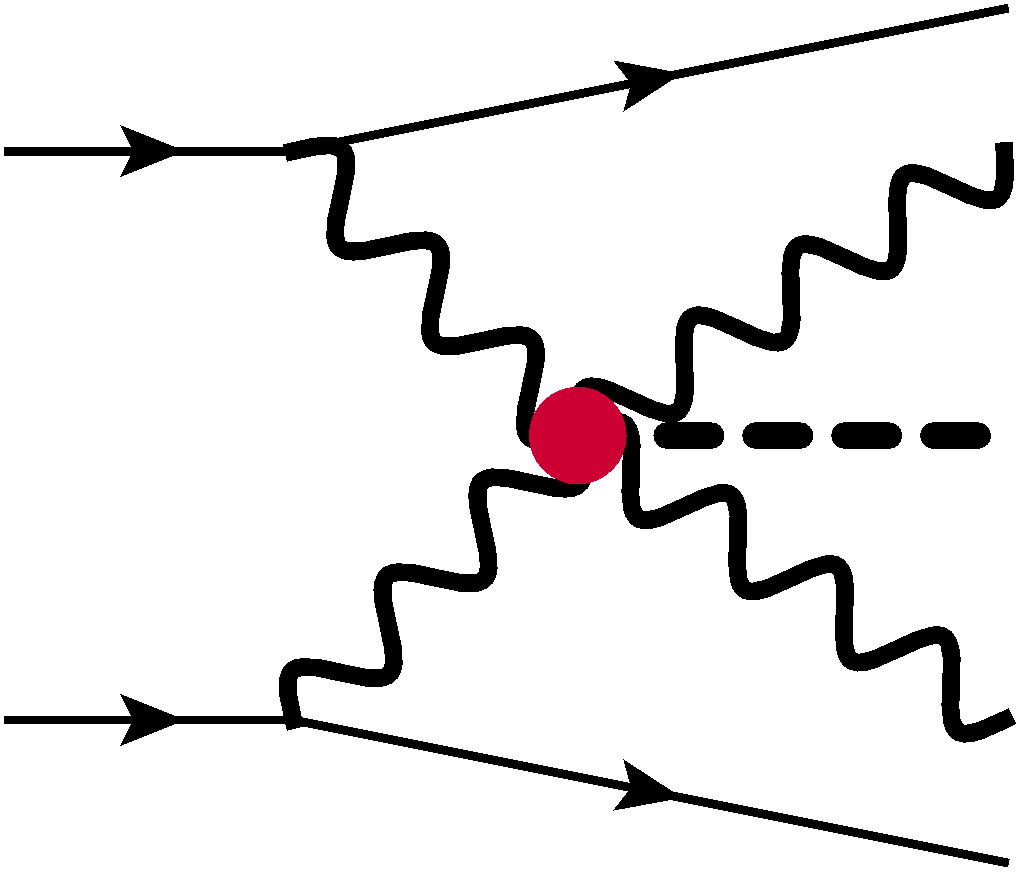}
			}} &$\sim\frac{vE}{\Lambda^2}$\\ \cline{1-5}
			{$\begin{array}{lll} \kappa_{Z\gamma}\\ \kappa_{\gamma\gamma}\\\kappa_{V}\end{array}$} &$\begin{array}{lll} \op_{WW}\\ \op_{BB}\\\op_{r}  \end{array}$ & {\resizebox{1.6cm}{1.2cm}{
					\includegraphics[width=0.3\textwidth]{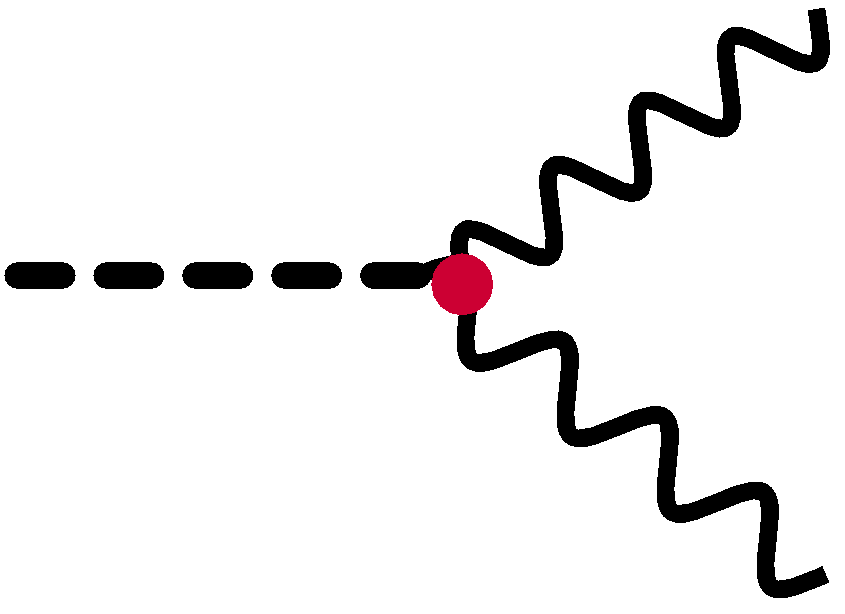}
			}} & {\resizebox{2cm}{1.8cm}{
					\includegraphics[width=0.3\textwidth]{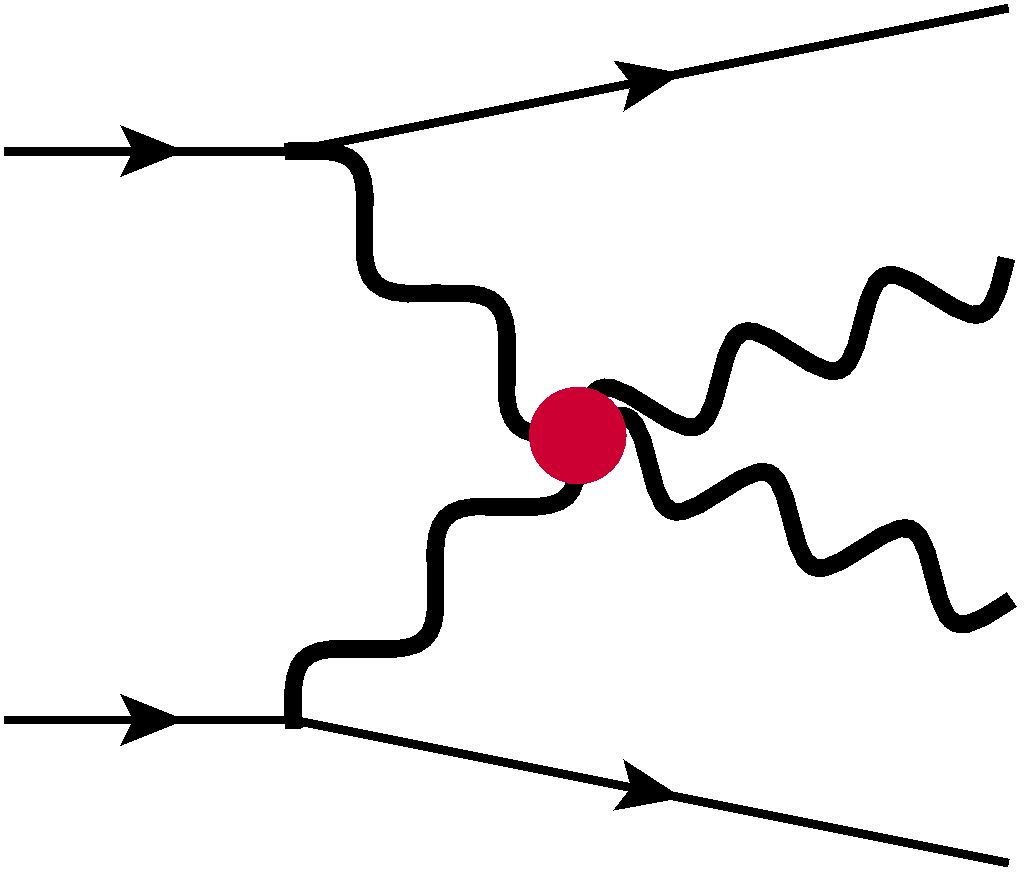}
			}} &$\sim\frac{E^2}{\Lambda^2}$\\ \cline{1-5}
			{$\kappa_g$} &{$\op_{gg}$} & {\resizebox{1.6cm}{1.2cm}{%
					\includegraphics[width=0.3\textwidth]{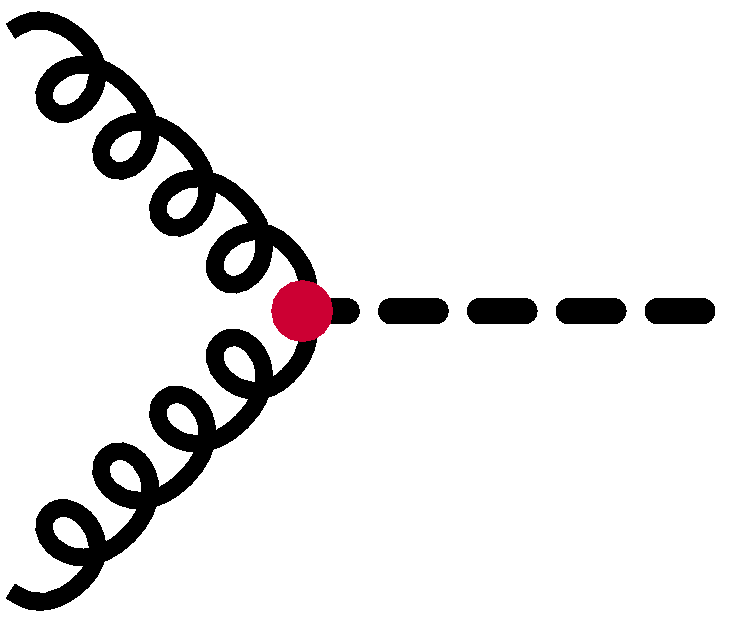}
			}} & {\resizebox{1.8cm}{1.2cm}{%
					\includegraphics[width=0.3\textwidth]{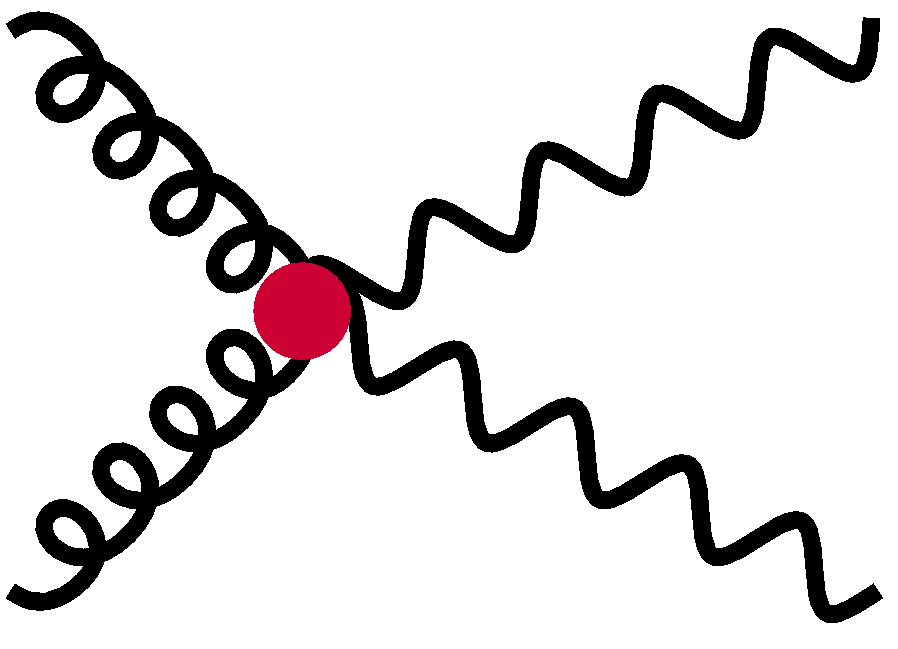}
			}}&$\sim\frac{E^2}{\Lambda^2}$\\ %\cline{1-3}
		\end{tabular}
	\end{center}
	\caption{\footnotesize \emph{Each effect (left column) can be measured as an on-shell Higgs Coupling (diagram in the HC column) or in a high-energy process (diagram in the HwH column), where it grows with energy as indicated in the last column.% \com{put also $E^2$ and $E$ to highlight the E-growth and make it readable from the first page directly? Order according to text.}
	}}\label{tab:proc}
\end{table} 

In this letter we initiate a novel program to test the very same Higgs couplings, \emph{off-shell} and at high-energy, via their contributions to the physics of longitudinally polarized gauge bosons. We will show that this program is potentially competitive with  \emph{on-shell} measurements. %, while additionally offering endless opportunities for refinements and improvements. 
%With ``high-energy'' we mean kinematic regions where systematics errors are of the same order as statistical ones; for this reason
Moreover---and perhaps equally important---this program contains numerous avenues for refinements and improvements: 
%Indeed, the high-energy program
it can benefit maximally from accumulated statistics, from improved SM computations of differential distributions, from  phenomenological analyses aimed at enhancing the signal-over-background (see, for instance, \cite{Farina:2016rws,Franceschini:2017xkh,Alioli:2017jdo,Azatov:2017kzw,Panico:2017frx,Banerjee:2018bio}), and  from dedicated experimental analyses. %aimed at reducing the different backgrounds.
Furthermore, given the complexity of the final states, we expect advanced machine learning techniques \cite{Brehmer:2018kdj,Christensen:2008py,DAgnolo:2018cun} could drastically improve our simple cut and count analysis. 
Additionally, in the context of a global precision program, the high-energy aspects that we discuss here will be the ones that benefit the most, not only from the long-term HL-LHC program, but also from potential future high energy colliders, such as the High-Energy (HE) LHC or CLIC.

Our leitmotiv is that \emph{any} observable modification of a SM coupling will produce in \emph{some} process a growth with energy (see table~\ref{tab:proc}). In some sense, this is obvious: since the SM is the only theory that can be extrapolated to arbitrarily\footnote{Modulo the Landau pole and the coupling to gravity, both irrelevant for the present discussion.} high-energy, any departure from it can have only a finite range of validity, a fact that is made manifest by a disproportionate growth in certain scattering amplitudes. Theories with a finite range of validity are, by definition, EFTs; for this reason the best vehicle to communicate our message is  the EFT language of \eq{eq:ops}. We stress nevertheless that at, tree level, the very same conclusions can be reached in the $\kappa$ framework \cite{Heinemeyer:2013tqa} or in the unitary-gauge framework of Ref.~\cite{deFlorian:2016spz,Gupta:2014rxa}.

The operators of \eq{eq:ops} have the form $|H|^2\times \op^{\mathrm SM}$, with $\op^{\mathrm SM}$ a dimension-4  SM operator (\textit{i.e.} kinetic terms, Higgs potential, and Yukawas) times
\begin{equation}\label{h2}
|H|^2=\frac{1}{2}\left(v^2+ 2 hv + h^2+2\phi^+\phi^-+(\phi^0)^2\right)
\end{equation}
where $v=246$ GeV is the Higgs vacuum expectation value (vev), $h$ is the physical Higgs boson, and $\phi^{\pm,0}$ are the would-be longitudinal polarizations of $W$- and $Z$-bosons.
From the operators in \eq{eq:ops}, the piece $\propto v^2$ can be reabsorbed via a redefinition of the SM input parameters and is therefore unobservable \cite{Pomarol:2013zra,Elias-Miro:2013mua}; the piece $\propto v h$ constitutes instead the core  of the HC measurements program, as it implies modifications to single-Higgs processes (triple Higgs processes for $\op_6$), and can be matched easily to the $\kappa$ framework. 
The $h^2$ piece was discussed in~\cite{Goertz:2014qta,Bishara:2016kjn,Azatov:2015oxa} {in the context of double Higgs production}.
In this article we focus on the last two terms in \eq{h2} and study processes with longitudinal gauge bosons instead of processes with an on-shell Higgs; we dub this search strategy ``Higgs without Higgs''  - HwH in short.  

The high-energy avenue is potentially very promising: for $E^2$-growing effects, a 1\% sensitivity at the Higgs boson mass, corresponds to a $O(1)$ sensitivity at $E\sim 1$ TeV.
We will see that, in practice, High-$E$ measurements are rather complex, so that this na\"ive scaling is hardly achieved in the explorative analysis presented here. However, we envisage several strategies for improvement that outline a
 challenging and exciting collider program. 
 
 %We expose our philosophy in section \ref{sec:HEproc}, where we identify the relevant high energy processes associated with each Higgs coupling, and estimate in an exploratory way their collider reach. In \textcolor{red}{BH: this needs to finish}

 \section{High-Energy Processes}\label{sec:HEproc}

The first ingredient in this program is to identify which processes  grow maximally with energy once  Higgs Couplings are modified. There is a quick and intuitive way of to assess this based on \emph{1)} dimensional analysis, \emph{2)}  our choice of EFT basis \eq{eq:ops}, and \emph{3)}  on the parametrization chosen in  \eq{h2}, where the longitudinal polarizations are explicitly represented by their scalar high-energy counterpart~\cite{Cornwall:1974km,Chanowitz:1985hj,Wulzer:2013mza}. For $v\to0$, the operators of \eq{eq:ops} contribute directly to contact interactions with $n=4$ fields ($\op_{WW},\op_{BB},\op_{GG},\op_r$), 5 fields ($\op_{y_\psi}$) or 6 fields ($\op_H$), with a coupling $\propto 1/\Lambda^2$ that carries two inverse powers of mass dimensions. Amplitudes generated by these contact vertices do not involve any propagators (which carry inverse powers of energy) and are therefore maximally energy-growing. %\footnote{This reasoning is a peculiarity of the Warsaw basis\cite{Grzadkowski:2010es}; in the SILH basis, for instance,  some operators are proportional to the equations of motion and derivatives at denominator cancel the propagator in the amplitude...unitary gauge...\com{could remove this ftnt}} 
%{\color{red} Recalling that amplitudes with $n$ legs have dimension $[\mathcal{A}_n]=\GeV^{4-n}$}
At high-energy---$E\gg m_W,m_h,m_t$---the only other dimensionful parameter is the energy $E$; hence, generically, we expect that the BSM and SM contributions to the same process scale as
\begin{equation}
\frac{\mathcal{A}^{\op}_n}{\mathcal{A}^{\textrm{SM}}_n}\sim \frac{E^{2}}{\Lambda^2}\,.
\end{equation}

\begin{figure}[t]
	\begin{center}
\includegraphics[width=0.3\textwidth]{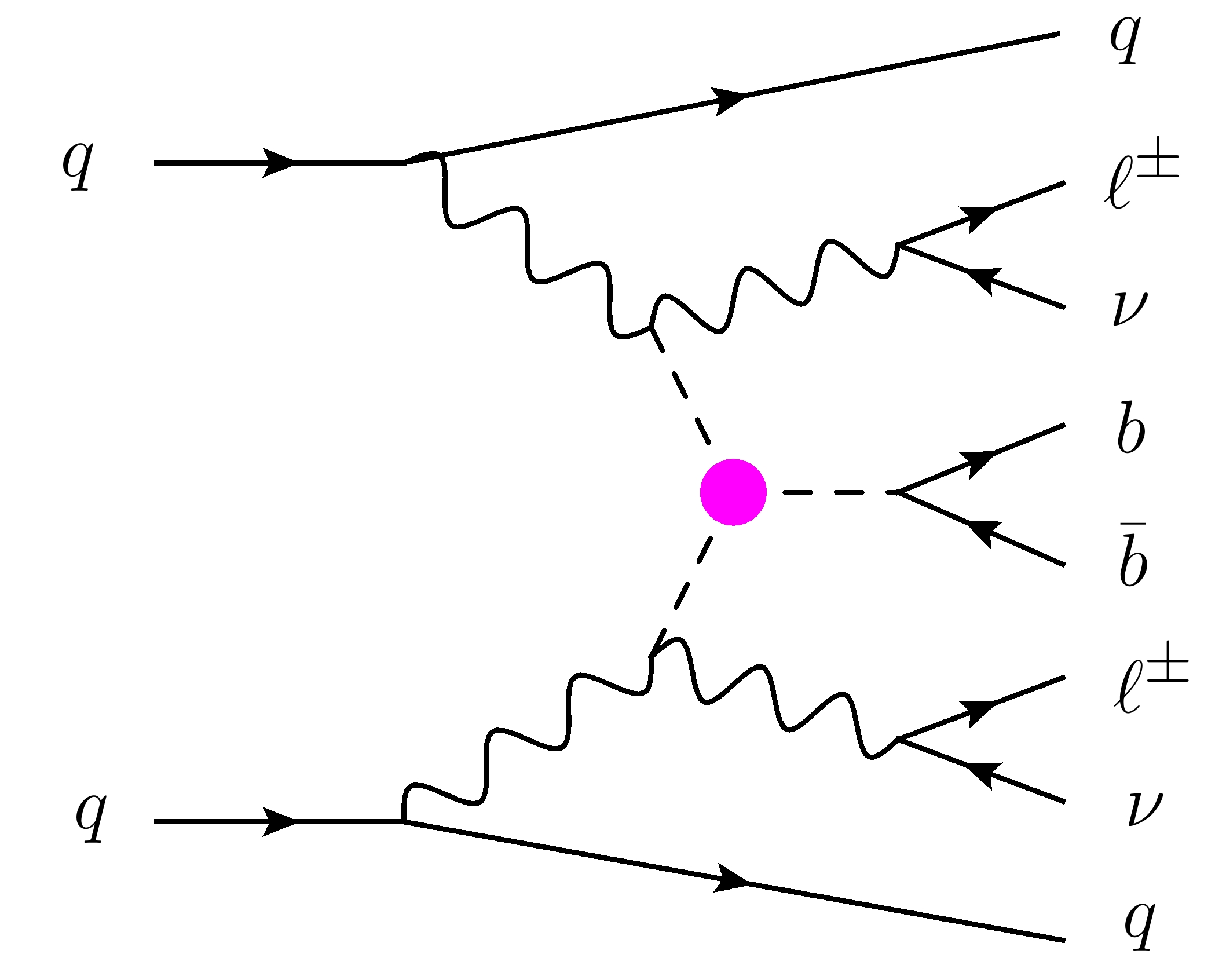}
	\end{center}
	\caption{\footnotesize \emph{
            An energy-growing process sensitive to the Higgs trilinear, \eq{prockh3}. Here we show the diagram in unitary gauge; it is equivalent to the one in Table~\ref{tab:proc} where the Goldstones are kept explicit. 
	    %A unitary-gauge diagram with energy-growing sensitive to the Higgs trilinear.
            The two VBF jets and, in particular, same-sign leptons give rise to an exceptionally clean channel.}}\label{fig:unitary}
\end{figure} 

Table \ref{tab:proc} shows the relevant processes that exhibit this behaviour; more explicitly, at hadron (lepton) colliders, 
\begin{eqnarray}
&&\kappa_t: p p \to j t+ V_LV^\prime_L\label{prockt}\\
&&\hspace{.6cm}(e^+e^-\to ll+\{tbW_L,tbZ_L,ttW_L,ttZ_L\})\nn\\
&&\kappa_\lambda:  p p \to  j j h + V_LV^\prime_L, \,\, (e^+e^-\to llhV_LV^\prime_L)  \label{prockh3}\\
 &&\hspace{.7cm}p p \to jj+4V_L,\,\, (e^+e^-\to ll\,4V_L) \label{prockh36}\\
 &&\kappa_{\gamma\gamma,Z\gamma}:  p p \to j j +V^\prime V,\,\,\, (e^+e^-\to ll V^\prime V)\label{prockga} \\ 
&&\kappa_V:  p p \to jj+V_LV^\prime_L,\,\,\,(e^+e^-\to llV_LV^\prime_L)\label{prockv}\\
&&\kappa_g:  p p \to W_L^+W_L^-, Z_LZ_L ,\,\,\,(e^+e^-\to ll j j ) \label{prockg}
\end{eqnarray} 
where $V_LV^\prime_L\equiv\{W_L^\pm W_L^\pm,W_L^\pm W_L^\mp,W_L^\pm Z_L,Z_LZ_L\}$ (similarly $4V_L$ a generic longitudinally polarized final state) and $V^{(\prime)}$ any (longitudinal or transverse) vector, including photons, while $l$ denotes either a charged lepton $\ell^\pm$ or a neutrino,  depending on the final state.
To help visualize our discussion in terms of HC, Fig.~\ref{fig:unitary} shows a diagram exhibiting $E$-growth in unitary-gauge. 
%We also show in Fig. 1 a unitary-gauge
%diagram that exhibits E-growth and helps visualize
%our discussion in terms of HC.
%Notice that, for all processes, the  amplitude associated with the modified couplings grows quadratically with the relevant energy scale of the process $E^2$ (with the exception of \eq{prockh3}, see later).
Notice that the  amplitudes associated with the modified couplings grow quadratically with energy $E^2$ (with the exception of \eq{prockh3}, see later).

In the following paragraphs we explore these processes in turn and provide a first estimate of the potential HwH reach at the HL-LHC in comparison with the reach from Higgs couplings measurements. %This rhetoric of competitiveness has the sole scope of providing the reader with a quantitative feeling about the power of HwH processes; it is understood that, for practical purposes, the two search methods should be thought of as complementary. 
Our results are based on leading order (LO) MadGraph simulations \cite{Alwall:2014hca}, where the Higgs couplings have been modified using FeynRules \cite{Christensen:2008py} and checked against the model of Ref. \cite{Falkowski:2015wza}.
%Before concluding we comment on the extension of this program to future colliders.

%\begin{widetext}
\begin{figure}[ht]
\begin{center}
\includegraphics[width=0.3\textwidth]{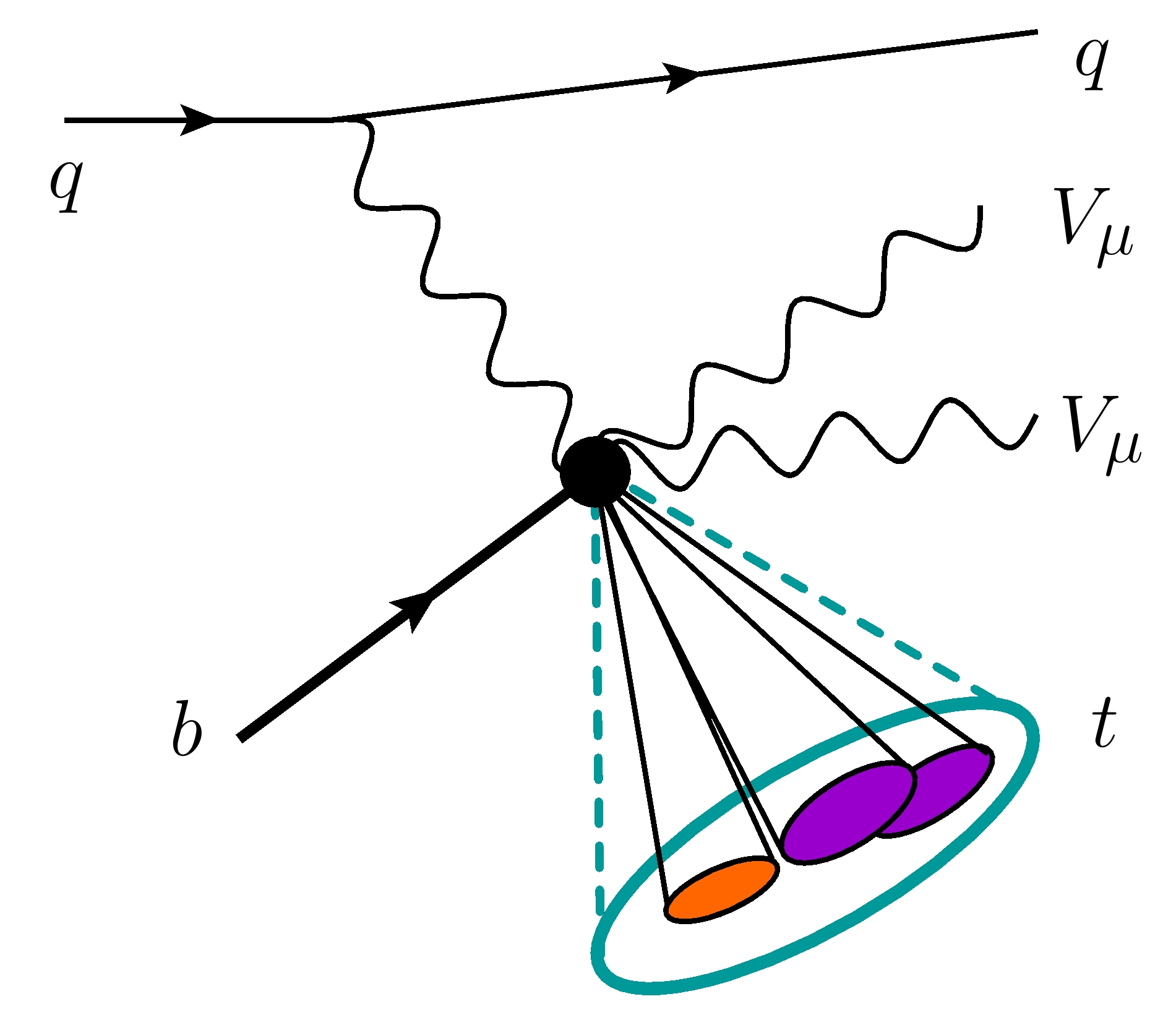}
\end{center}
     \caption{\footnotesize \emph{Process sensitive to the top Yukawa, \eq{prockt}. The boosted single top and the forward jet tag the event. The analysis is binned in the number of leptons, from the vector boson decays.}}\label{fig:top}
\end{figure} 

 %\begin{widetext}
\begin{figure*}[t]
\begin{center}
\includegraphics[width=0.32\textwidth]{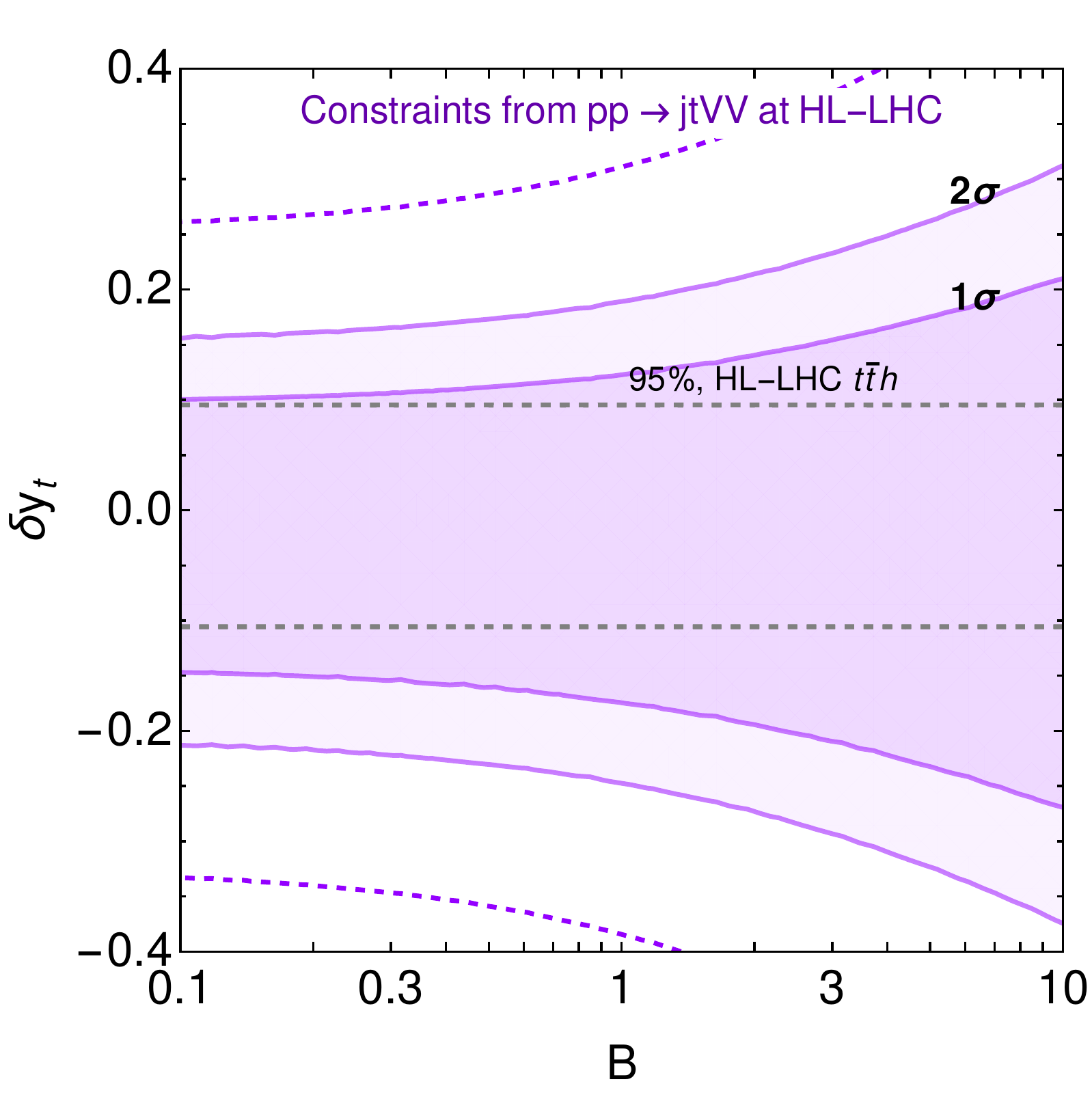}
\includegraphics[width=0.32\textwidth]{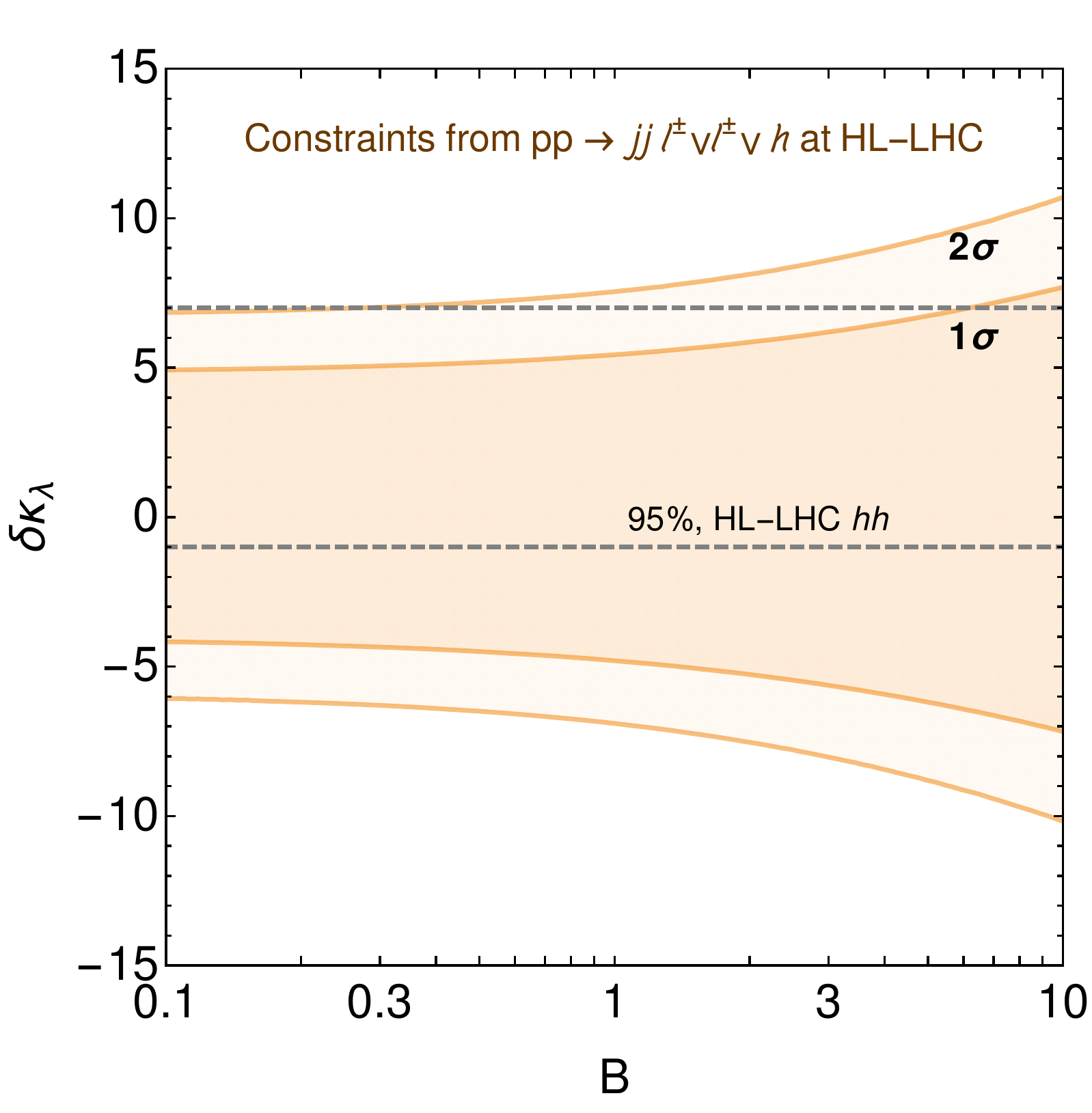}
\includegraphics[width=0.30\textwidth]{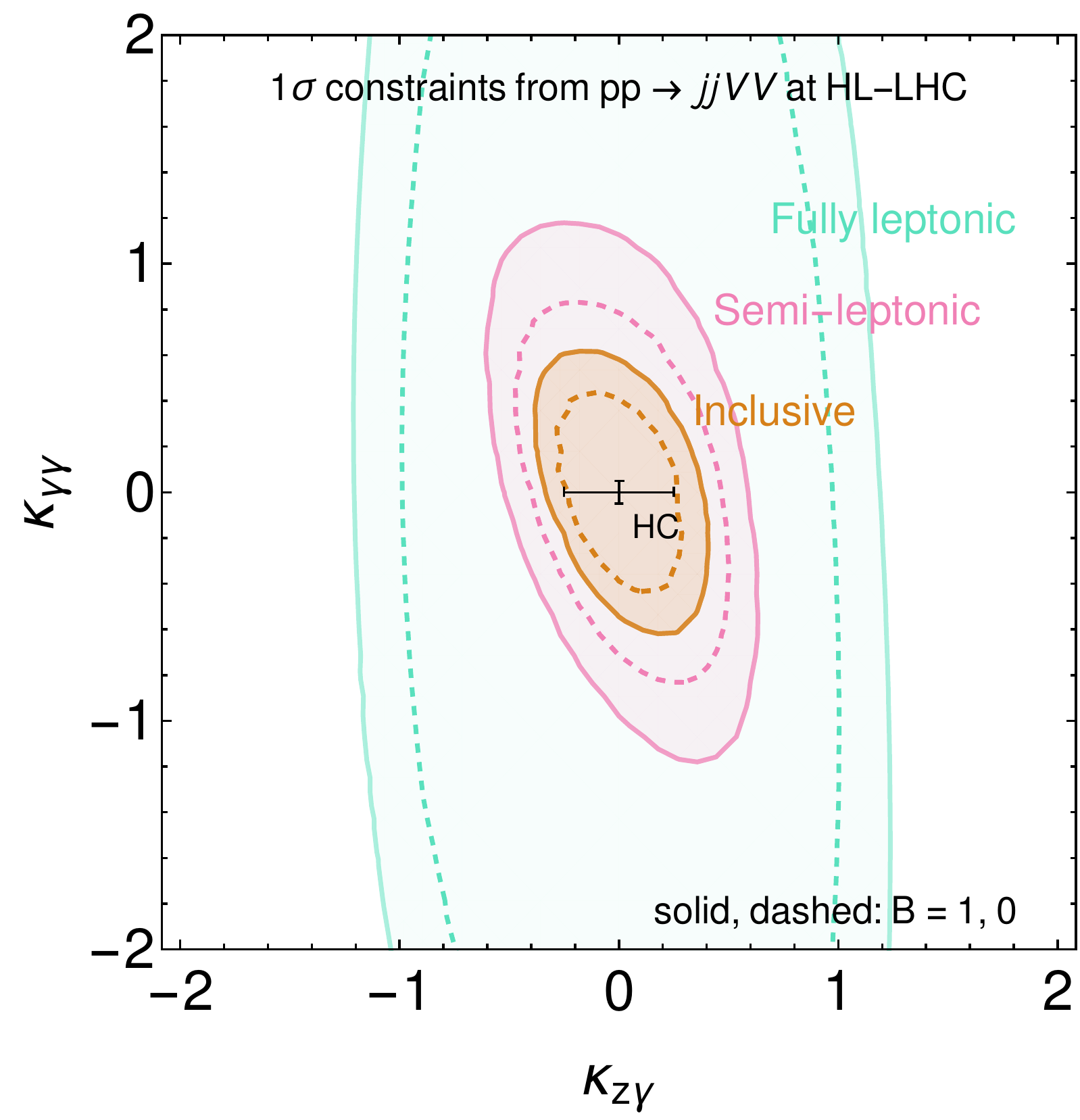}
\end{center}
     \caption{\footnotesize \emph{LEFT: HL-LHC (3000 fb$^{-1}$) sensitivity on modifications of the top quark Yukawa $\delta y_t$ from the process in Fig.~\ref{fig:top} (shaded bands), and from measurements of Higgs couplings (95\%C.L., dashed grey lines); $B$ controls additional backgrounds (for $B=1$ the analysis includes a number of background events equal to the SM signal); the dashed purple line corresponds to $1\sigma$ results for the $\ge 2\ell$ categories (which have the smallest backgrounds). CENTER: same, but for modifications of the Higgs trilinear $\delta\kappa_\lambda$. RIGHT: $1\sigma$ reach for modification of the Higgs-$\gamma\gamma$ and $Z\gamma$ rates, using high-$E$ measurements (green,pink,brown bands correspond to leptonic,semileptonic, and also hadronic final states)  or Higgs couplings (black error bars). 
}}\label{fig:reach}
\end{figure*} 
%\end{widetext}

\vspace{5mm}
\noindent
{\bf The top Yukawa.}
Modifications of the Yukawa coupling of the Higgs boson to top quarks is reputedly difficult to measure on the $h$ resonance~\cite{Aaboud:2018urx}; however, an anomalous top quark Yukawa induces a quadratic energy growth in the five point amplitude $\mathcal{A}(bV_L \to tV_LV_L)$ involving a bottom quark, a top, and three longitudinal bosons. {This amplitude leads to a process with a final state consisting of a top quark, a forward jet and two longitudinally polarized vector bosons, see \eq{prockt} and Fig.~\ref{fig:top}. Notice that these have a smaller energy threshold compared with the $tth$ final state used in HC measurements.\footnote{See also Ref.~\cite{Degrande:2018fog} that  studies $thj$ final states which exhibits linear $E$-growth with modifications of the top-Yukawa.}

  The top carries a large transverse momentum $p_T^t$ due to the hardness of the process, which makes it a good discriminator. We consider two categories, for $p_T^t>250(500)$~GeV. A forward jet with $|\eta_j|>2.5$, $p_T^j>30$~GeV and $E_j>300$ GeV is required.

  The signal is classified by  counting the number of extra leptons reconstructed in the event. The following table shows the number of signal events  at the 14 TeV HL-LHC  with 3000 fb$^{-1}$, for $p_T^t>250/500$~GeV,
  
\vspace{.2cm}

\hspace{-3mm}\begin{tabular}{|c|c|c|c|c|c|}
\hline
Process& 0$\ell$ & 1$\ell$ & $\ell^\pm\ell^\mp$ & $\ell^\pm\ell^\pm$ & $3\ell(4\ell)$\\
 \hline
 $W^\pm W^\mp$    &  3449/567 & 1724/283 & 216/35 & - & -  \\
 $W^\pm W^\pm$    &  2850/398 & 1425/199 & - & 178/25 & -  \\
 $W^\pm Z$   &  3860/632& 965/158 & 273/45 & - & 68/11 \\
 $Z Z$    &  2484/364& - & 351/49 & - &  (12/2) \\\hline
\end{tabular}\vspace{.2cm}

The categories with two or more leptons have small background. For the hadronic modes---which dominate the 0 and 1 lepton channels---the largest source of background comes from $\bar t t jj\to tWbjj$ where the $b$ quark gets misidentified as an ordinary jet and the two lowest rapidity jets reconstruct a $W/Z$-boson. After applying the event topology selection cuts---the required forward jet, the invariant mass of the two lower rapidity jets reconstructs an EW gauge boson mass, and a boosted top---the cross-section is 470 fb (22 fb) for $p_T^t > 250$ GeV ($> 500$ GeV), roughly 80 (20) times that of the signal. However, in order to fall into the signal region, the $b$ quark must be misidentified as a regular jet and the pair of lower rapidity jets must mimic a hadronically decaying vector. The $b$ misidentification rate is order 10\% for a 90\% light jet acceptance~\cite{ATL-PHYS-PUB-2017-013}. Vector boson tagging techniques~\cite{ATLAS-CONF-2018-016} can identify a hadronically decaying vector with a $10^2$ background rejection for a 40\% signal efficiency. A conservative estimate of the combined effect of these cuts brings the background to comparable or smaller size than the signal.

We broadly parametrize this and other backgrounds by a uniform rescaling $B$ of the SM signal expectation in each bin (so that for $B=1$ we add an irreducible background equal to the SM signal in each channel), and show the estimated reach in the left panel of Fig.~\ref{fig:reach}. The dashed grey lines compare our results with those from HC measurements~\cite{Cepeda:2019klc}. For illustration we also show  results that focus on channels with at least 2 leptons with a dashed purple line: here the backgrounds are much smaller. The large number of events left in the zero and one lepton categories makes it possible to extend the analysis to higher energies, where not only the effects of the energy growth will be enhanced, but also the background reduced; a dedicated study is needed to assess more precisely the acceptances of these hadronic channels.

This mode of exploration also appears well-suited for high-energy lepton colliders like CLIC. Indeed, the processes in the second line of \eq{prockt} have a lower threshold for production than the $\bar tth$ final state that is usually considered to measure the top quark Yukawa. Moreover, the final state in \eq{prockt} is produced in vector boson fusion, whose crossection increases with energy, while $\bar tth$ production proceeds via Drell-Yan, which decreases with energy.% We plan to study this in detail in the future.

\vspace{5mm}
\noindent
{\bf The Higgs self coupling.}
Measurements of the Higgs self-coupling have received enormous attention in collider studies.  In the di-Higgs channel at HL-LHC precision can reach $\delta\kappa_\lambda\in[-1.8,6.7]$ at 95\%C.L.~\cite{ATL-PHYS-PUB-2017-001} using the $b\bar{b}\gamma\gamma$ final state. 
Here we propose the processes of Eqs.~(\ref{prockh3},\ref{prockh36}) with VBS scattering topology and a  multitude of longitudinally polarized vector bosons, see the  second row of Tab.~\ref{tab:proc} as well as Fig.~\ref{fig:unitary}. 
The modified coupling $\delta \kappa_\lambda$, or the operator $\op_6$, induces a linear growth  with energy w.r.t. the SM in processes with $jj h V_LV_L $ final state (Tab.~\ref{tab:proc}), and a quadratic growth in processes with  $jj V_LV_LV_LV_L$. For the former, the same-sign $W^\pm W^\pm h jj$  with leptonic $(e,\mu)$ decays is particularly favourable for its low background: two same-sign leptons (2ssl) and VBS topology offers a good discriminator against background, allowing for $h\to\bar b b$ decays. For illustration we focus on this channel in which the SM gives $N_{\textrm{SM}}\simeq 50$ events. {Backgrounds from $ t \bar t jj$ enter with a mis-identified lepton, but it can be shown that they can be kept under control with the efficiencies reported in \cite{Khachatryan:2015hwa} and with VBS cuts on the forward jets. A potentially larger background is expected to come from fake leptons, but the precise estimation is left for future work.}

The results---shown in the center panel of Fig.~\ref{fig:reach}---are very encouraging: this simple analysis can match the precision of the by-now very elaborate di-Higgs studies.
There are many directions in which this approach can be further refined: \emph{i)} including the  many other final states  in \eq{prockh3}, both for the vector decays and for the Higgs decay \emph{ii)} including the  $E^2$-growing $jj V_LV_LV_LV_L$ topologies  of \eq{prockh36}, \emph{iii)} taking into account differential information.
Moreover, the process of Tab.~\ref{tab:proc} grows only linearly with energy w.r.t. the SM amplitude with transverse vectors in the final state, but it grows quadratically w.r.t. the SM longitudinal final states, so: \emph{iv)} measurements of the polarization fraction can improve this measurement. %We leave all this for a future detailed study.

\vspace{5mm}
\noindent
{\bf Higgs to $\gamma\gamma,Z\gamma$.}
These decay rates are loop-level and small in the SM: their measurement implies therefore  tight constraints on possible large (tree-level) BSM effects, which in the EFT language are captured by the operators $\op_{WW,BB}$ from \eq{eq:ops}.\footnote{The same operators also affect the  $h$ couplings to $Z_TZ_T$ and $W_TW_T$. The same qualitative analysis can be performed with focus on the $hA_{\mu\nu}A^{\mu\nu}$ and  $hA_{\mu\nu}Z^{\mu\nu}$ vertices, but we prefer to work here with the gauge invariant $\op_{WW,BB}$ operators. See also comments in section \ref{sec:CC}.}
These also enter in high-energy VBS \eq{prockga}, and they represent a beautiful additional motivation (together with $\kappa_V$, see below) to study these processes, which at present are often interpreted in the context of anomalous quartic gauge couplings (QGC)~\cite{Eboli:2006wa}, corresponding to dimension-8 operators.

We perform a simple analysis of vector boson scattering (VBS) with $W^\pm W^\pm, ZZ, WZ,Z\gamma$ final states. For the first three we cut as usual on the forward jets: $|\delta_{jj}|>2.5$, $p_T^j>30$~GeV and $m_{jj}>500$~GeV \cite{Aad:2014zda}. A kinematic variable that captures the hardness of the $2\to 2$ process is the scalar sum of the $p_T^V$ of the vector bosons, and therefore we divide the distribution in bins of 250 GeV up to 2 TeV. For the $Z\gamma$ final state, we follow the analysis for aQGC of \cite{Aaboud:2017pds}.

The combined results are displayed in the right panel of Fig.~\ref{fig:reach}, for fully leptonic, semileptonic and fully hadronic decays  (a difficult challenge at the LHC), for backgrounds  $B=0,1$ where, as explained above, $B=1$ corresponds to an additional background of the same order as the SM. Note that we translated the constraints on $c_{BB},c_{WW}$ to the $\kappa_{\gamma\gamma},\kappa_{z\gamma}$. We find that the $ZZ,Z\gamma$ final states provide the best reach.
For comparison, the individual reach from HL-LHC measurements of HC \cite{Cepeda:2019klc} is shown by the black error bars. These clearly offer an unbeatable sensitivity in the $h\gamma\gamma$ direction; the $hZ\gamma$ direction is however less tested, and our simple analysis of high-energy probes shows promising results. Notice that the $2\to 2$ amplitudes generated by $\kappa_{\gamma\gamma},\kappa_{z\gamma}$ are of the form $V_L V_L\to V_TV_T$ (and permutations thereof) and do not interfere in inclusive measurements~\cite{Azatov:2016sqh}; this fact might provide a hint on how to make these analyses perform better.

\vspace{5mm}
\noindent
{\bf Higgs to $W^+W^-,ZZ$.}
It is  known that modifications of the tree-level $hZZ$ and $hW^+W^-$ SM couplings (assumed here to be controlled by a unique parameter, corresponding for instance to $\op_H$ in the SILH basis \cite{Giudice:2007fh}) imply a quadratic $E$-growth in longitudinal VBS. This is discussed in detail in Ref.~\cite{Contino:2010mh} (and \cite{Contino:2013gna} for linear colliders), where it is pointed out that, in the SM, the longitudinal component is suppressed by an accidental factor $\sim 2000$, which is equivalent to a very large irreducible background. This motivated studies of VBS $hh$ pair production instead, see \cite{Bishara:2016kjn}, finding at $1\sigma$, $\delta\kappa_V\lesssim 8\%$, comparable to $\delta\kappa_V\lesssim 5\%$ from HC \cite{Cepeda:2019klc}.\footnote{The authors of \cite{Bishara:2016kjn} assume separate couplings of the vector bosons to $h$ or $h^2$; when the Higgs is part of a doublet, these are proportional. Moreover, the numbers we report here are indicative: both HC measurements and the di-higgs analysis have optimistic and pessimistic scenarios in which these numbers might differ.}

%%We believe, however, that a combination of diboson channels ...better?
%\com{polarization tagging, 100 TeV? Contino does 1\%...couplings? }

\vspace{5mm}
\noindent
{\bf Higgs to $gg$.} %\com{remove plot and comment only?}
This coupling modifies the main production mode at hadron colliders and is, therefore, very well measured. %\com{check}: as such it provides an interesting territory to compare high-energy versus high-luminosity observables. 
The most interesting high-energy process that can be associated with this coupling is $gg\to ZZ$, which has  been discussed in Refs.~\cite{Azatov:2014jga,Cacciapaglia:2014rla,Azatov:2016xik}. 
Using the results from Ref.~\cite{Azatov:2014jga} we estimate HwH versus HC reach at the end of the HL-LHC, %in particular we have considered a scenario with and one without the background and three different decay channels . We find that 
\begin{eqnarray}
&\textrm{HC:}\,\,\,\, & |\delta\kappa_g|\lesssim 0.025\nn\\
&\textrm{HwH:}&|\delta\kappa_g| \lesssim 0.24\, /\, 0.06 \, /\, 0.01\\
&\textrm{HwH}&(\textrm{no }\bar qq\to Z_TZ_T):\,\, |\delta\kappa_g| \lesssim 0.09  \, /\, 0.02  \, /\, 0.005\nn
\end{eqnarray}
where the numbers stand for the fully leptonic / semileptonic / fully hadronic channels. 
%\begin{align}
%\textrm{HC}\qquad\qquad\quad\ \ & \\
%\qquad\qquad \kappa_g\lesssim 0.025&\\
%\textrm{HwH}\qquad\qquad\quad& \\
%\textrm{no }\bar qq\to Z_TZ_T,\ \kappa_g \text{ in } & :\\
% [-0.087,0.093]&  \text{, Fully Leptonic}  \\
% [-0.021,0.025]&  \text{, Semi Leptonic}  \\
% [-0.005,0.005]&  \text{, Fully Hadronic} \\
%\textrm{with }\bar qq\to Z_TZ_T,\ \kappa_g \text{ in }&:\\
% [-0.20,0.24]&  \text{, Fully Leptonic}  \\
% [-0.055,0.058]& \text{, Semi Leptonic}  \\
% [-0.014,0.014]&  \text{, Fully Hadronic} \\
%\end{align}
The partonic $\bar qq\to Z_TZ_T$ process represents here the main irreducible background, as it does not interfere with our $gg\to Z_LZ_L$  amplitude with longitudinal polarization. Its reduction would constitute an important aspect of HwH analyses. Notice that, unfortunately and similar to the $\kappa_{\gamma\gamma},\kappa_{z\gamma}$ cases discussed above, the SM-BSM interference is also suppressed because of the different helicities~\cite{Azatov:2016sqh}. Despite these difficulties---which might be overcome in more refined analyses (along the lines of \cite{Panico:2017frx,Azatov:2017kzw,Dixon:1993xd})---the high-$E$ results remain competitive  in the semileptonic and fully hadronic channels, assuming that the background from $\bar qq\to Z_TZ_T$ can be efficiently suppressed.

%Despite these difficulties, which might be overcome in more refined analyses (along the lines of \cite{Panico:2017frx,Azatov:2017kzw}), the high-$E$ results remain competitive  in the semileptonic and fully hadronic channels, assuming that the background from $\bar qq\to Z_TZ_T$ can be efficiently suppressed. 

The amplitude we propose can also find a beautiful implementation in the context of future lepton colliders, in the form of $ZZ,WW\to gg$ in VBS. There, the possibility to polarize the initial electron positron beams could offer an additional handle to enhance the longitudinal polarizations.
 This would offer a new potential for ILC or CLIC to improve upon Higgs coupling measurements.

%
%
%
%\section{Future Colliders}\label{sec:future}
%
%gradual improvement
%
%
%
%
%A generic advantage of high-energy probes is that
%
%
%
%
%
%\emph{ii)} interference \emph{resurrection} (along the lines of \cite{Panico:2017frx})\footnote{The largest amplitude in the SM has $TT$ polarization, while our effect is $LL$; $LL-TT$ interference can be very small in practice, because of the difficulty of distinguishing left- vs right- handed leptons \cite{Panico:2017frx}. It would be nice to explore this possibility also in the context of lepton colliders in the VBS $ZZ\to gg$ process, where the initial beam can be polarized.}
%

\section{Comments}\label{sec:CC}

In a generic EFT fit there can be other operators that enter in the observables  we propose. On general grounds, assuming the Higgs is part of a doublet, %\footnote{We comment here that our use of the EFT operators in Eq.~(1) assumes the Higgs boson belongs to an electroweak doublet, as is suggested by existing measurements of the Higgs� couplings. More generally, this need not be the case; one can work with the HEFT. In such a case, the channels we propose correspond to operators involving longitudinal vectors and are not necessarily related directly to Higgs couplings. This possibility merits further study; nevertheless, we emphasize that the channels we study here are still important tests of BSM physics as encoded by the HEFT.}
 we expect these operators to be better constrained by other measurements, so that their impact on our study is small. 
Dividing operators by the number of fields they contain, equivalent to the number of legs $n$ of the first amplitude to which they contribute as $\sim E^2/\Lambda^2$, we have~\cite{Grzadkowski:2010es},
\begin{center}
\begin{tabular}{|l|l|l|}
\hline
n&HC & No HC \\
\hline
$6$ &$h^6 $ & \\
$5$ &$\psi^2h^3 $ & \\
$4$ &$ F^2h^2,h^4 D^2 $& $F\psi^2 h, \,\psi^4,\, \psi^2 h^2 D,\, h^4 D^2$ \\
$3$ & & $F^3$  \\
\hline
\end{tabular}
\end{center}
%\begin{gather}
%n=6:\, h^6 \quad \quad
%n=5:\, \psi^2h^3\quad\quad n=3:\, F^2 \\
%n=4:\, F^2h^2,\, F\psi^2 h, \,\psi^4,\, \psi^2 h^2 D,\, h^4 D^2
%\end{gather}

\noindent where $F,\psi,h$ denote field strengths, spinor and scalar (Higgs) fields, and $D$ denotes derivatives, and we have divided operators that modify HC from those that do not.
In the latter category, the majority contributes to $2\to 2$ partonic processes, where they are expected to be better measured than in processes with more legs.\footnote{Non-interference arguments \cite{Azatov:2016sqh} can lead to processes with more  legs   providing better measurements \cite{Dixon:1993xd,Azatov:2017kzw}. For electroweak processes, however, this is superseded by measurements of azimuthal angles  from the decay of vector bosons~\cite{Azatov:2017kzw,Panico:2017frx}.}
So we  expect our studies of the $n=6,5$ processes in Eqs.~(\ref{prockt},\ref{prockh3})---targeting  $\kappa_{t}\sim \psi^2h^3$ and $\kappa_{\lambda}\sim h^6$---to be rather robust against the presence of other operators.

On the same lines, in our analysis of $gg\to ZZ$ targeting $\kappa_{g}\sim \op_{GG}$, there are no other dimension-6 operators that enter with $E^2$-growing effects; this would not have been true for the    $W^+W^-$ final state (also modified by $\op_{GG}$).

The structure $h^4D^2$  appears in two combinations (e.g. $\op_{HD}$, $\op_{H\Box}$ in the Warsaw basis). The combination that cannot be associated with Higgs couplings enters in the  $T$ parameter; this is very  well measured from LEP~\cite{Peskin:1991sw,Barbieri:2004qk}, so that we can associate $V_LV_L$ scattering almost entirely with modifications of the Higgs coupling to vectors $\kappa_V\sim \op_r$.

For the other effects that enter VBS, the situation is more complex and requires a detailed study. Nevertheless, the majority of $n=4$ operators that do not modify HC contain fermions (and $W_{\mu\nu}^3$  modifies amplitudes $A(\psi\psi\to VV^\prime)$ involving fermions), so that these operators can be measured also in other processes. For this reason  we  expect that HwH observables will remain important also in the context of a global fit.

We also mention that  if the Higgs is not part of a doublet (i.e. a non-linear realisation of electroweak symmetry with a singlet, also known as HEFT), then contributions to the processes in Eqs.~(\ref{prockt}-\ref{prockg}) from operators involving only the longitudinal polarizations, will be decorrelated from the operators involving only the physical Higgs entering HC processes, thus potentially offering an opportunity to distinguish between HEFT and SM EFT.

 %\footnote{We comment here that our use of the EFT operators in Eq.~(1) assumes the Higgs boson belongs to an electroweak doublet, as is suggested by existing measurements of the Higgs� couplings. More generally, this need not be the case; one can work with the HEFT. In such a case, the channels we propose correspond to operators involving longitudinal vectors and are not necessarily related directly to Higgs couplings. This possibility merits further study; nevertheless, we emphasize that the channels we study here are still important tests of BSM physics as encoded by the HEFT.}

\vspace{5mm}

An important aspect  of the high-$E$ exploration is the question of EFT validity, in this respect we comment as follows. 
Any deviations from the SM predictions of the processes considered imply an ultimate cutoff of the theory: the scale of unitarity violation $\Lambda_{sc}$, where the particles of the SM would become strongly coupled in the absence of new dynamics. The possibility of a consistent EFT interpretation requires $E < \Lambda_{sc}$, see \textit{e.g.}~\cite{Biekoetter:2014jwa,Contino:2016jqw,Pobbe:2017wrj}. $\Lambda_{sc}$ depends on the size of the deviations; for instance, for $O(1)$ deviations from the Higgs  self-coupling $\kappa_\lambda$ this scale is of order 13 TeV~\cite{Chang:2019vez,Falkowski:2019tft}, much larger than the typical relevant energy that we are accessing at the~LHC. Other Higgs couplings are better measured and hence have larger strong-coupling scales associated, and therefore a wider EFT validity range. These considerations imply that our analysis---which utilizes high-energy bins of differential distributions---admit a consistent EFT treatment.

It is also important to consider the validity of truncating the EFT expansion when a measurement is performed in regions with poor sensitivity, in a way that it relies on the quadratic $O(1/\Lambda^4)$ terms in the cross section. In light of this there are different ways in which our analysis can be taken: \emph{i)} we believe it will be possible to refine and redesign our analyses, along the lines of \cite{Farina:2016rws,Franceschini:2017xkh,Alioli:2017jdo,Azatov:2017kzw,Panico:2017frx,Banerjee:2018bio} (see also comments in the $hgg$ paragraph), in order to make it more precise and sensitive to the linear terms  $O(1/\Lambda^2)$ in the expansion, \emph{ii)} in strongly coupled theories the inclusion of $O(1/\Lambda^4)$ does not imply a breakdown of the expansion \cite{Biekoetter:2014jwa,Contino:2016jqw,Pobbe:2017wrj}, so that unrefined high-energy measurements can always be interpreted in this context, \emph{iii)} a breakdown of the EFT implies generically the discovery of on-shell modes within the kinematic collider reach; given that the sensitivity we find here is comparable to that of HC, our analysis implies that the question of EFT validity should be discussed also in the context of HC measurements.

\section{Conclusions}\label{sec:Conc}
%\vspace{5mm}
%\noindent
%{\bf Comments and Conclusions.}
In this article we have proposed a novel way of testing Higgs couplings at colliders, based on high-energy processes rather than processes involving an on-shell Higgs  resonance.
Exploiting the fact that anomalous modifications of the SM necessarily induce \(E\)-growth in some process, we identified and initiated an explorative study of the potential reach of these high-energy probes, which we have compared with Higgs coupling measurements in the context of the HL-LHC.
%We  exploit the fact that anomalous modifications of the SM necessarily introduce energy-growth in some process, which we have identified, and initiated an explorative study of the potential reach of these high energy probes, which we have compared with Higgs coupling measurements in the context of the HL-LHC.

The preliminary results are very positive,
 especially given the potential of improvements that we foresee.  
Simple cut-and-count analyses were shown, in some cases, to match the precision of sophisticated Higgs Coupling measurements.
For instance, the $jjW^\pm W^\pm h$  channel with leptonic decays, allows a precision comparable to di-Higgs production in measuring the Higgs self-coupling. Similarly, modifications of the top Yukawa can be measured in  the many $j t+ V_LV^\prime_L$ final states to a precision in the ballpark of Higgs coupling measurements.
VBS processes and $ZZ$ at high-energy offer further alternative possibilities to test the Higgs coupling to electroweak gauge bosons and to gluons, respectively.

There are many directions in which our analysis can be extended and refined,  and which we will investigate in future work.
\emph{1)} Realistic estimates of the relevant backgrounds and acceptances in the different channels, are the first step in the HwH program; this motivates the development of tools, along the lines of Ref.~\cite{ATLAS-CONF-2018-016}, to reject QCD background from a hadronically decaying vector-boson signal.
\emph{2)} Moreover, our signals center on the presence of longitudinally polarized vector bosons. An important irreducible background in this context is the SM fraction of transversely polarized states which, in many cases,  is much larger than the longitudinal signal~\cite{Contino:2010mh}.
So, important progress could come from a better understanding of the kinematics of the various helicity amplitudes, aimed at improving the signal/background ratio, along the lines of~\cite{Dixon:1993xd,Panico:2017frx,Azatov:2017kzw}, or other techniques that access information about vector-boson polarization~\cite{CMS:2018zxa,CMS:2018mbt,Lee:2018xtt}.
\emph{3)}~A more refined understanding of the relevant scales and BSM-sensitive distributions in the problem, as in~\cite{Degrande:2018fog}, or complemented with more advanced BDT or machine learning techniques, see e.g.~\cite{Brehmer:2018kdj,Christensen:2008py,DAgnolo:2018cun}, would also increase HwH sensitivity. \emph{4)}~A realistic analysis would also include QCD corrections; these tend to increase the relevant cross sections, but also complicate their helicity structure. \emph{5)} Finally, it would be interesting to include our observables in the context of a global fit (see e.g. \cite{Pomarol:2013zra,Gomez-Ambrosio:2018pnl,Biekotter:2018rhp,deBlas:2016ojx,Ellis:2018gqa}): we expect that, even in situations where they cannot directly compete with individual HC measurements, they can still provide valuable global information by potentially lifting flat directions. %\emph{6)} Finally, if the Higgs is not part of a doublet or EWSB effects are large (i.e. a non-linear realisation of electroweak symmetry with a singlet, also known as HEFT), then contributions to the processes in Eqs.~(\ref{prockt}-\ref{prockg}) from operators involving only the longitudinal polarizations, will be decorrelated from the operators involving only the physical Higgs entering HC processes, thus potentially offering an opportunity to distinguish between HEFT and~SM~EFT.

%%%%%%%%%%%%%%%%%%%%%%%%%%%%%%%%%
\subsection*{Acknowledgements}
We thank  Pietro Govoni, Matthew McCullough, Johnny Raine and Steven Schramm for valuable discussions. This project is funded by the Swiss National Science Foundation under grant no. PP002-170578.
\bibliography{Bib.bib}

\end{document}